\documentclass[10pt,conference,a4paper]{IEEEtran}

\def\vh{{\vect{h}}} 
\def\vx{\vect{x}}
\def\vb{\vect{b}}

\def\mb{\matr{B}}
\def\train{^{\circ}}
\def\test{}

\def\sampx{x_{i}}
\def\sampvb{{\vb}_{i}}
\def\sampb{b_{i}}

\def\varn{\sigma_{w}^{2}}
\def\invvarn{\sigma_{w}^{-2}}

\newcommand{\mean}[2]{\overline{#1}_{#2}}
\newcommand{\mvar}[1]{C_{#1}}

\newcommand{\likevxb}[3]{p(\vx{#1}|\vb{#2},{#3})} 
\newcommand{\likexbh}[1]{p(x_{i}{#1}|\sampvb{#1},\vh)}

\newcommand{\posthD}[1]{p({#1}|\tset)} 

\newcommand{\evidvx}[2]{p(\vx{#1}|{#2})} 
\newcommand{\evidxbdes}[2]{p(x_{1}{#1},\ldots,x_{#2}{#1}|b_{1}{#1},\ldots,b_{#2}{#1})}

\newcommand{\priorvh}{p(\vh)}

\newcommand{\probbx}[1]{p(b_{i}{\test}=b|x_{1}\test,\ldots,x_{N}\test,#1)}
\newcommand{\probbvxval}[2]{p(b_{i}{\test}={#2}|\vx{\test},#1)}

\newcommand{\varkay}{(\id+\mb\train(\mb\train)^{\top}\invvarn)^{-1}}

\newcommand{\convmatrix}[3]{\mb_{#2}{#1}=[{\vb}_{1}{#1},{\vb}_{2}{#1},\ldots,{\vb}_{#3}{#1}]}

\newcommand{\distribas}{\thicksim}

\newcommand{\vect}[1]{\mathbf{#1}}     
\newcommand{\matr}[1]{\mathbf{#1}}     

\def\bve{{\mathbf \bve}}

\def\tset{\mathcal{D}}

\def\Nor{\mathcal{N}}                

\newcommand{\id}{\matr{I}}       

\reversemarginpar

\usepackage{amsmath}
\usepackage{amsfonts}%
\usepackage{amssymb}

\usepackage{graphicx}%
\DeclareGraphicsExtensions{.png,.eps,.ps,.pdf}
\begin{document}

\title{Channel Decoding with a Bayesian Equalizer}

\author{
\authorblockN{Luis Salamanca, Juan Jos{\'e} Murillo-Fuentes}
\authorblockA{DTSC
University of Seville,\\
Camino de los Descubrimientos s/n,\\
41092 Seville, Spain.\\
{\tt \{salamanca,murillo\}@us.es}}
\and
\authorblockN{Fernando P{\'e}rez-Cruz}
\authorblockA{DTSC
University Carlos III in Madrid,\\
Avda. de la Universidad 30, 28911,\\
Legan{\'e}s (Madrid), Spain.\\
{\tt fernando@tsc.uc3m.es}
}
}

\maketitle

\begin{abstract}
Low-density parity-check (LPDC) decoders assume the channel estate information (CSI) is known and they have the true a posteriori probability (APP) for each transmitted bit. But in most cases of interest, the CSI needs to be estimated with the help of a short training sequence and the LDPC decoder has to decode the received word using faulty APP estimates. In this paper, we study the uncertainty in the CSI estimate and how it affects the bit error rate (BER) output by the LDPC decoder. To improve these APP estimates, we propose a Bayesian equalizer that takes into consideration not only the uncertainty due to the noise in the channel, but also the uncertainty in the CSI estimate, reducing the BER after the LDPC decoder.
\end{abstract}

\section{Introduction}

Single input single output (SISO) communication channels can be characterized by a linear finite impulsive response that either represents the dispersive nature of a physical media or the multiple paths of wireless communications \cite{Proakis08}. This representation causes inter-symbol interference (ISI) at the receiver end that can impair the digital communication. Given this channel estate information (CSI), the maximum likelihood sequence detector (MLSD) \cite{Viterbi73} -Viterbi Algorithm- provides the optimal transmitted sequence at the receiver end. And, if we are interested in a posteriori probabilities (APP) for the transmitted symbols, we can use the BCJR algorithm \cite{Bahl74}, that provides bitwise optimal decisions. 

Channel encoders introduce controlled redundancy in the transmitted sequence to correct errors caused by the channel. Modern channel decoders, such as turbo or low-density parity-check (LDPC) codes \cite{Urbanke08-2}, need accurate APP estimates to be able to achieve channel capacity \cite{Cover06}.

In a previous work, we have shown that accurate APP estimates increase the performance of LDPC decoders \cite{Olmos2010}, although that work focuses on nonlinear channel estimation.

In this paper, we study how the uncertainty in the estimation of the CSI affects the optimal performance of modern channel decoders. The CSI is acquired using a training sequence \cite{Proakis08}  and it is typically estimated by maximum likelihood (ML). These training sequences are necessary short to reduce the transmission of non-informative symbols, yielding inaccurate CSI estimates. Thus, the BCJR assuming an ML estimation (hereafter refers as ML-BCJR) only delivers an approximation to the APP for each symbol because it does not include the uncertainty in the estimate. Inaccuracies in the APP estimates degrade the performance of channel decoders for turbo or LDPC codes \cite{Luby01,Chung01}. We consecutively propose and analyze a Bayesian equalizer, which takes into account the uncertainty in the CSI estimate to produce more accurate APP estimates.

The difference between the bit error rate (BER) of the ML-BCJR and the proposed Bayesian equalizer is not significant, although it slightly favors the Bayesian equalizer. However, this is not an accurate measure of the quality of the APP estimates for each equalizer since it only considers hard decisions, in contrast to the soft inputs needed by modern channel decoders. Thus, assuming LDPC coding in our communication system, we can compare the quality of the APP estimates for each equalizer. We experimentally show that at the output of the LDPC decoder the Bayesian equalizer considerably improves the performance of the ML-BCJR equalizer, when we measure the probability of error. These gains are more significant for high signal to noise ratios, channels with long impulsive responses and/or short training sequences.

 There are some related works where uncertainties are exploited on the estimation of the transmitted symbols. In the framework of turbo-receivers \cite{douillard95}, some approaches can be found in the literature that incorporate these uncertainties in the iterative process of equalization and decoding. It has been exploited in \cite{davis01}, where the authors use an MMSE to estimate the channel, and in \cite{otnes03,wang01}, where they do not focus on the optimal estimation of the APP. In \cite{lu02} we find a proposal to estimate some parameters in a OFDM system to later include them in the decoding. In \cite{Piantanida09}, the authors consider the channel estimation inaccuracies during the decoding process by means of a practical decoding metric.

The paper is organized as follows. In Section \ref{sec2} we describe the structure of a SISO communication system and the ML-BCJR solution. The proposed Bayesian equalization technique is presented in Section \ref{sec3}.  Experimental results in Section \ref{sec4} help to illustrate the performance of our method. Finally, Section \ref{sec5} ends with conclusions and some proposals for future work.

\section{ML-BCJR Equalization}\label{sec2}
\begin{figure*}[!t]
\centering
\includegraphics[width=14.5cm]{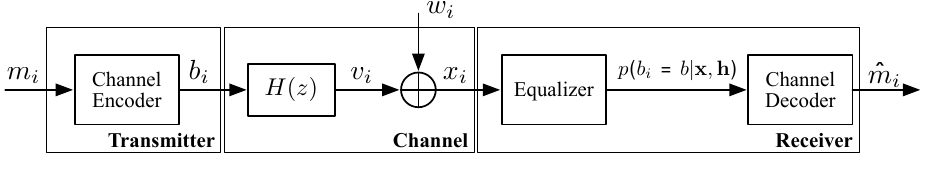}
\caption{System model.}\label{sistemaCanal}
\end{figure*}

We consider the discrete-time dispersive communication system depicted in Fig. \ref{sistemaCanal}. The channel $H(z)$ is completely specified by the CSI, i.e., $\vh=[h_{1},h_{2},\ldots,h_{L}]^{\top}$, where $L$ is the length of the channel. We model the values of the channel $\vh$ as independent Gaussians with zero-mean and variance equal to $1/L$ (Rayleigh fading).

A block of $K$ message bits, $\mathbf{m}=[m_{1},m_{2},\ldots,m_{K}]^{\top}$, is encoded with a rate $R=K/N$ to obtain the codeword $\vb=[b_{1},b_{2},\ldots,b_{N}]^{\top}$ that is transmitted over the channel using a BPSK modulation:
\begin{equation}\label{chtrans}
\sampx=\sampvb^{\top}\vh+w_{i},
\end{equation}
where $\sampvb=[\sampb,b_{i-1},{\ldots},b_{i-L+1}]^{\top}$ and $w_{i}$ is additive white Gaussian noise (AWGN) with variance $\varn$. Thus, the received sequence is $\vx=[x_{1},x_{2},\ldots,x_{N}]^{\top}$.

At the beginning of every block we transmit a preamble with $n$ known bits ($b_1^\circ,\ldots, b_n^\circ$) and the receiver uses $\tset=\{\sampx\train,\sampb\train\}^{n}_{i=1}$, the training sequence, to estimate the channel. The ML criterion is widely considered for the task of estimation:
\begin{equation}\label{MLcrit}
\hat{\vh}_{ML}=\arg\max_{\vh} \likevxb{\train}{\train}{\vh}.
\end{equation}

Once we have estimated the channel coefficients with the preamble, we apply the BCJR algorithm to obtain the posterior probability estimates for each transmitted bit:
\begin{equation}\label{probML}
\probbvxval{\hat{\vh}_{ML}}{b}\qquad\,i=1,2\ldots{N}.
\end{equation}

Finally we decode the received word using the LDPC decoder to obtain a maximum a posteriori estimate for $m_{i}$.

\begin{figure}[!h]
\centering
\includegraphics[width=9cm]{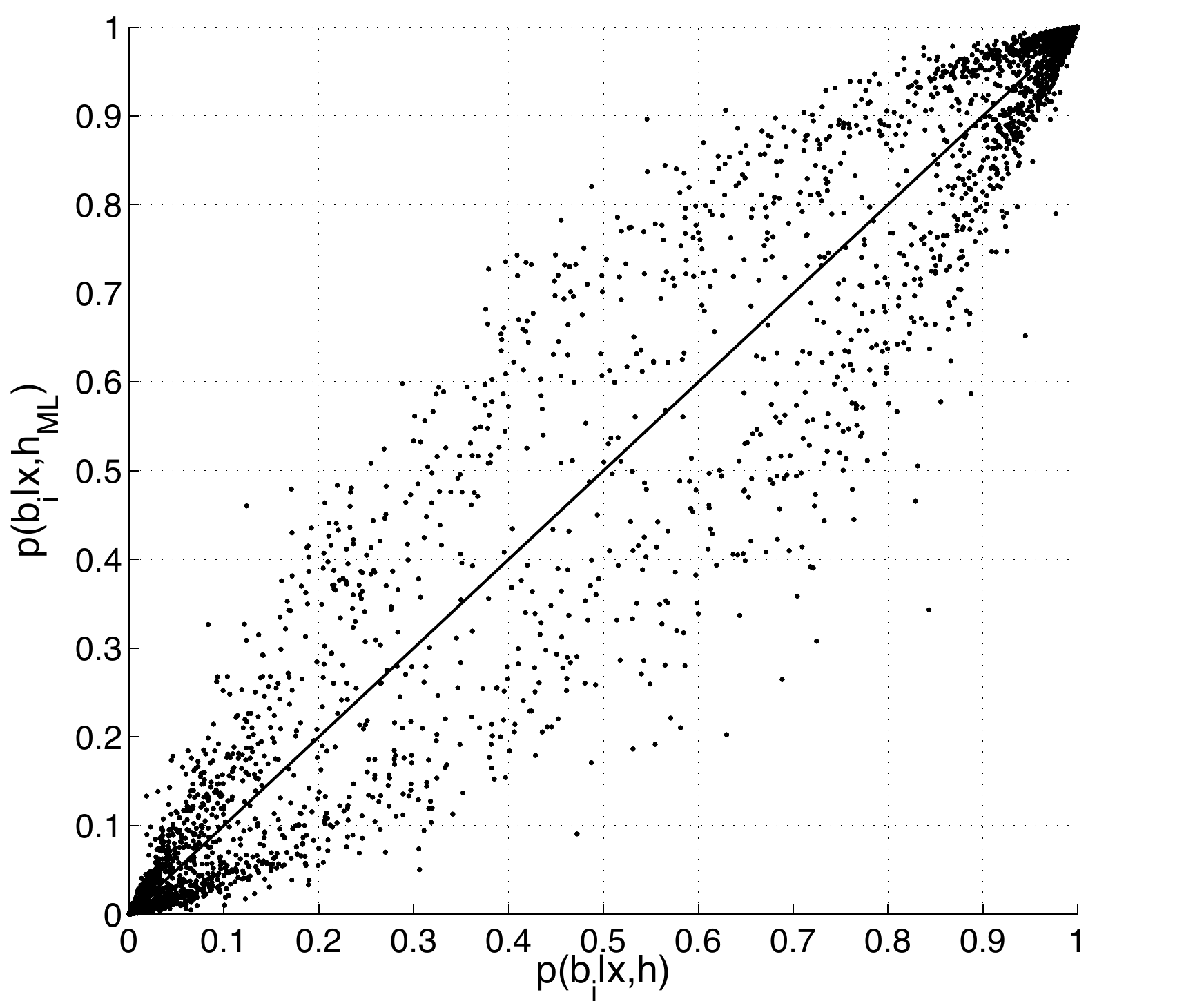}
\caption{Calibration curve for the ML-BCJR assuming $\mbox{SNR}=0$ dB and $\vh=[1,0.1]$.}\label{calcurvML}
\end{figure}

The LDPC decoder relies on the estimates in (\ref{probML}) being accurate and, if they are not, the decoding might not finish or might return an incorrect codeword. In Fig. \ref{calcurvML} we show $\probbvxval{\hat{\vh}_{ML}}{1}$ versus the probability $\probbvxval{\vh}{1}$ for one estimation of the fading channel $\vh=[1, 0.1]$ obtained assuming $n=6$, and a  signal to noise ratio (SNR) equal to $0$ dB. We can see in Fig. \ref{calcurvML} that the predictions for each bit are quite accurate. But these posterior probability estimates are overconfident roughly half of the time, because ML is an unbiased estimator of the CSI, which could derail the LDPC decoder, because a bit with high confidence for a zero or a one is hard to overrule if it is incorrect.

\section{Bayesian Equalization}\label{sec3}
If we provide the BCJR with the true CSI we have the exact APP \cite{Bahl74}. However, in practice we usually use the ML criterion in (\ref{MLcrit}) to estimate the CSI from a training sequence. This estimated CSI, provided as ground truth to the BCJR algorithm, assigns inaccurate estimates of the APP and might mislead the channel decoder to deliver the incorrect codeword (or not to converge at all). We propose a Bayesian equalizer that takes into account both the uncertainty about the noise and the uncertainty about the CSI estimate. We compute the posterior probability as:
\begin{equation}
\probbvxval{\tset}{b}=\int\probbvxval{\vh}{b}\posthD{\vh}d\vh\label{bayesint},
\end{equation}
where $b=\pm{1}$ and
\begin{align}
\posthD{\vh}=&\frac{\priorvh\likevxb{\train}{\train}{\vh}}{\evidvx{\train}{\vb\train}}\nonumber\\
=&\frac{\priorvh\prod_{i=1}^{n}\likexbh{\train}}{\evidxbdes{\train}{n}},
\label{postbayes}
\end{align}
is the posterior probability for the CSI, given the likelihood (Gaussian noise) and the prior (e.g. Rayleigh fading).

The marginalization of $\vh$ in (\ref{bayesint}) provides equal or better APP estimates that the ML-BCJR, because it includes all the information of $\posthD{\vh}$. In fact, if the training sequence is large enough, this Gaussian posterior tends to a multidimensional delta centered at its mean, whose value tends to the real value of $\vh$, as $\hat{\vh}_{ML}$ does. However, the performance of the ML-BCJR is misled in case of uncertainty in the CSI. Conversely, the Bayesian equalizer in (\ref{bayesint}) considers the uncertainty in the estimation, providing more accurate APP estimates.

This computation of the APP does not yield a significant improvement in the BER after the equalizer, because for detection we only consider if the probability is lower or higher than $0.5$. This is illustrated in Section \ref{sec4}.

\begin{figure}[!h]
\centering
\includegraphics[width=9cm]{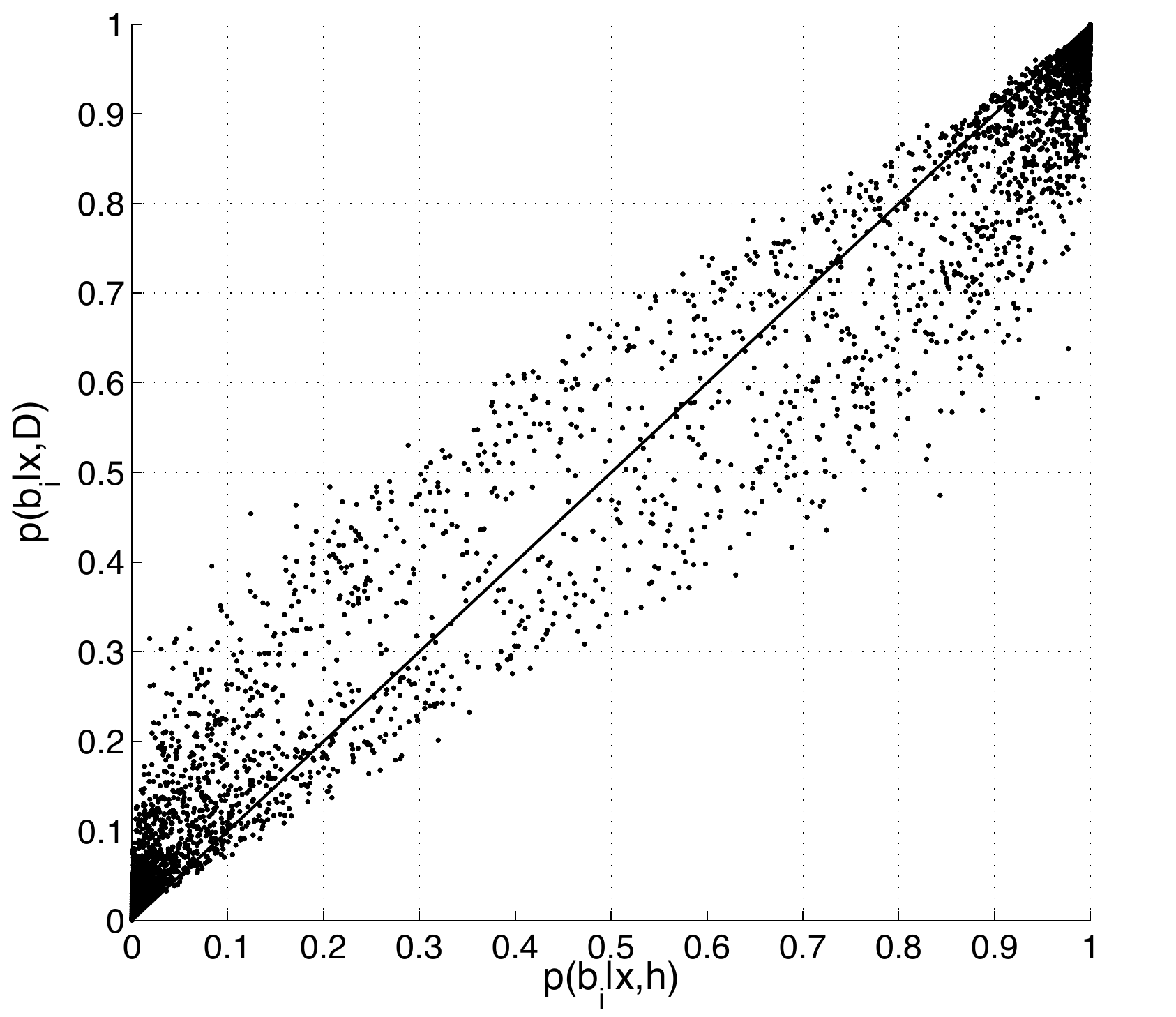}
\caption{Calibration curve for the Bayesian equalizer assuming $\mbox{SNR}=0$ dB and $\vh=[1,0.1]$.}\label{calcurvBay}
\end{figure}

However, the main advantage of the Bayesian equalizer is that the performance of the soft-decoder improves with this new estimation of the probabilities, that translates into better results in terms of BER at the output of the decoder. In Fig. \ref{calcurvBay} we show $\probbvxval{\tset}{1}$ versus $\probbvxval{\vh}{1}$ for a fading channel $\vh=[1,0.1]$ estimated with the same training sequences assumed in Fig \ref{calcurvML}. Compared to the ML-BCJR in Fig. \ref{calcurvML}, in the Bayesian equalizer the zero-mean Gaussian prior tends to underestimate the CSI. Then, the averaging over all possibles values of $\posthD{\vh}$ gives mostly underconfident posterior probability estimates, as we can observe in Fig. \ref{calcurvBay}. Therefore, the LDPC decoder does not have such a strong preference for the received bits and they can be easily flipped, if necessary.

\subsection{Computation of the solution}
We propose to use Monte Carlo to obtain an approximate result to (\ref{bayesint}) considering the following steps:

\begin{enumerate}
\item Calculate the posterior of the channel: in (\ref{postbayes}) the nu\-me\-ra\-tor is the product of the likelihood $\likevxb{\train}{\train}{\vh}$ and the prior of $\vh$. Considering a system with BPSK modulation in a Rayleigh channel,  both terms are real valued Gaussians distributed as:
\begin{align}
\likevxb{\train}{\train}{\vh}&\distribas\Nor(\,(\mb\train)^{\top}\vh,\varn\id)\label{distribpxhb},\\
\priorvh&\distribas\Nor(0,\id)\label{distribp},
\end{align}

where $\convmatrix{\train}{}{n}$ is an $L\times{n}$ matrix, and without loss of generality we assume that the variance of the CSI is equal to one. The expressions for the mean and the covariance matrix of the posterior when both terms are Gaussians are straightforward \cite{Kay93-1}. These expressions can be particularized for our system as:
\begin{align}\label{formpost}
\mean{\vh}{\vh|\tset}&=\varkay\mb\train\invvarn\vx\train,\\
\mvar{\vh|\tset}&=\varkay,
\end{align}
 
 where $\vx\train=[x_{1} \train,x_{2} \train,\ldots,x_{n} \train]^{\top}$.
\item Produce random samples from the posterior: with the vector of means and the covariance matrix, we can sample to obtain $M$ random samples.
\item Solve the BCJR algorithm: the APP estimates of each transmitted bit is computed for the $M$ different samples of $\posthD{\vh}$. 
\item Computation of (\ref{bayesint}): the $M$ different values of $\probbx{\vh}$ average the APP of each transmitted bit over all possible cases of $\vh$:
\begin{align}
&\probbx{\tset}\approx\nonumber\\
&\approx\frac{1}{M}\sum^{M}_{j=1}\probbx{\vh_{j}},
\end{align}
which yields an approximation of (\ref{bayesint}).
\end{enumerate}

As already discussed, if we use these optimal APP estimates as inputs to a soft-decoder, the system achieves better performance, especially in case of inaccurate estimations of the CSI, i.e., the length of the training sequence is short compared to the number of channel taps. However, this solution is computationally demanding, because we have to calculate $M$ times the BCJR algorithm, whose complexity increases exponentially with $L$. 

\section{Simulation Results}\label{sec4}
To illustrate the performance of the proposed receiver, we compare its bit error rate curves to the ones of the ML-BCJR, before and after the decoder. In all experiments we consider the following scenario:
\begin{itemize}
\item Block frames of 500 random bits encoded with a regular LDPC code (3,6) of rate $1/2$.
\item Up to 10000 frames of 1000 bits are transmitted over the channel.  
\item Between frames, a training sequence of $n$ random uncoded bits is transmitted to estimate the channel.
\item Every frame, and its associated training sequence, is sent over the same Rayleigh fading channel. We consider that the channel co\-he\-ren\-ce time is greater than the duration of the frame, i.e., the channel does not change during this time. Furthermore, in our experiments we take the same value for the taps of the channel during all transmitted frames. For the channel with three taps -$L=3$- these values are:
\begin{equation*}
H(z)=0.3482+0.8704z^{-1}+0.3482z^{-2},
\end{equation*}
and for $L=6$:
\begin{align*}
&H(z)=0.1600+0.5450z^{-1}-0.6720z^{-2}+\\
&+0.2560z^{-3}+0.0950z^{-4}-0.3890z^{-5}.
\end{align*}
\item We consider for the Bayesian estimation a prior with zero mean and variance equal to 1 for all the taps.
\end{itemize}

\begin{figure}[!h]
\centering
\includegraphics[width=9cm]{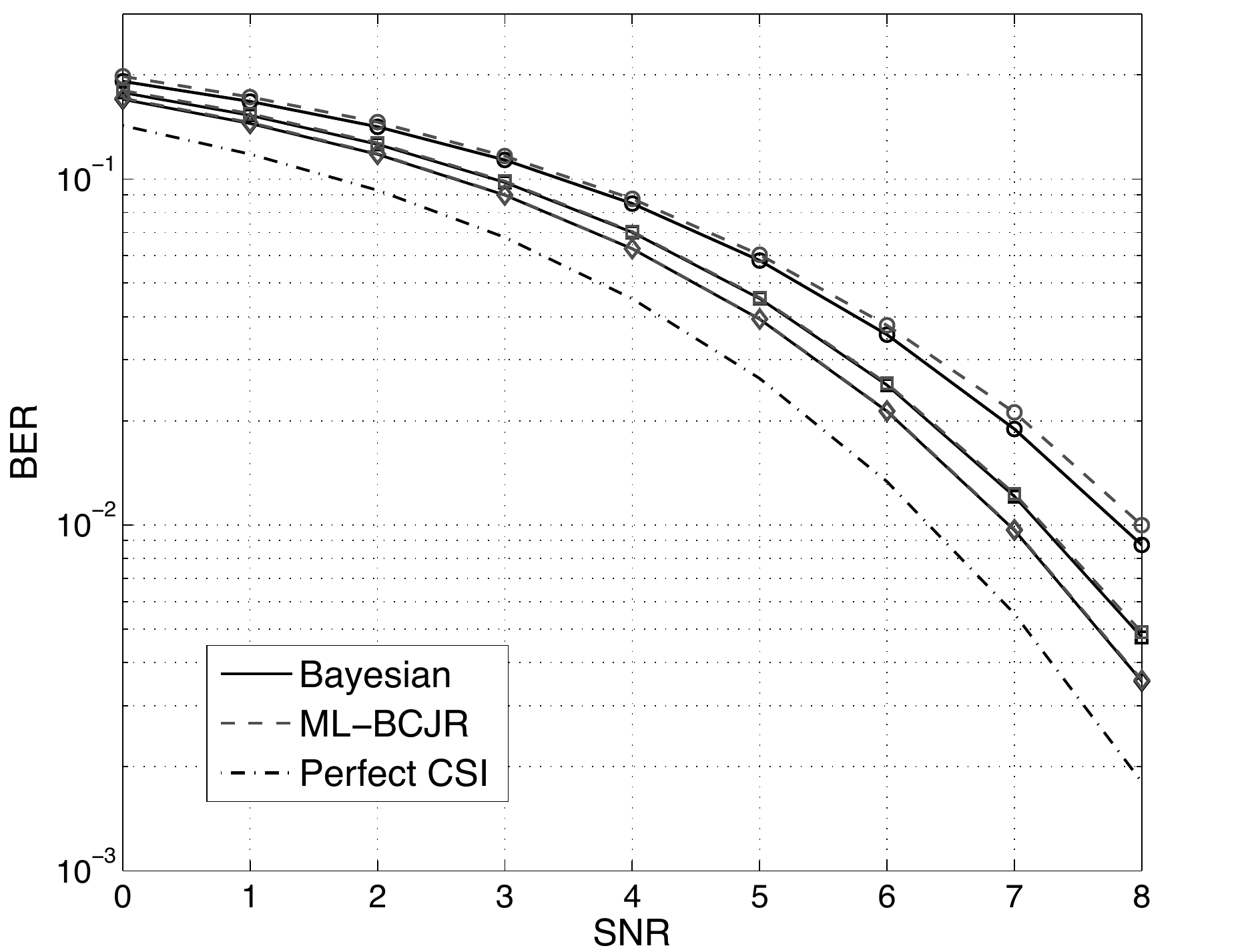}
\caption{BER performance for Bayesian equalizer (solid lines) and ML-BCJR (dashed lines) before the decoder, for a channel with 3 taps and different lengths of the training sequence, $n=10$ ($\circ$), $n=15$ ($\square$) and $n=20$ ($\diamond$). Dash-dotted line illustrates performance assuming perfect CSI.}\label{ldpcresul6tapsdetect}
\end{figure}

We depict in Fig. \ref{ldpcresul6tapsdetect} the BER measured after the ML-BCJR (dashed line) and Bayesian (solid line) equalizers for the channel with $L=3$ and different lengths of the training sequence. The differences between both equalizers are negligible, although the BER of the Bayesian equalizer is always lower.

\begin{figure}[!h]
\centering
\includegraphics[width=9.1cm]{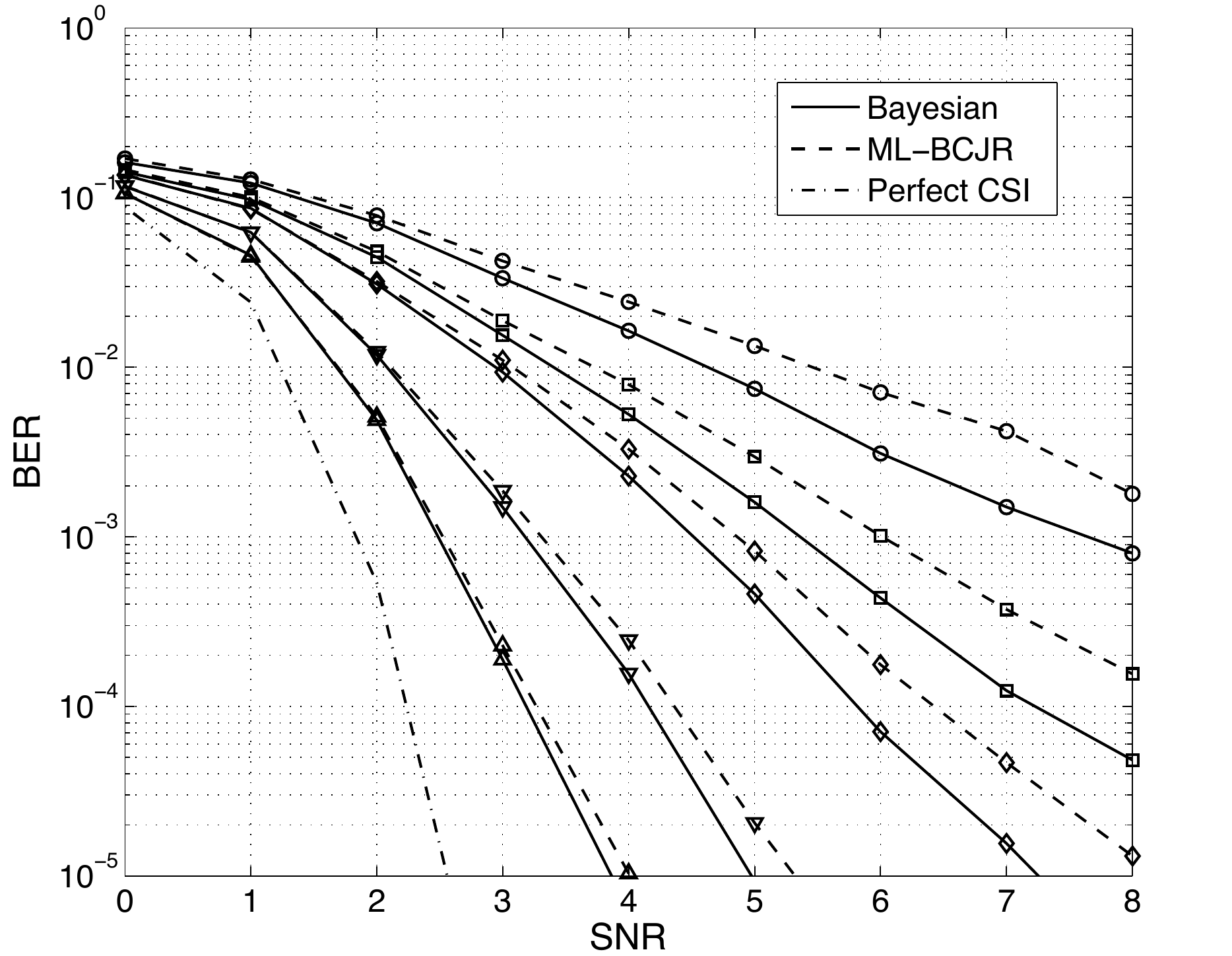}
\caption{BER performance for Bayesian equalizer (solid lines) and ML-BCJR (dashed lines) after the LDPC decoder, for a channel with 3 taps and different lengths of the training sequence, $n=10$ ($\circ$), $n=15$ ($\square$), $n=20$ ($\diamond$), $n=35$ ($\bigtriangledown$) and $n=60$ ($\bigtriangleup$). Dash-dotted line illustrates performance assuming perfect CSI.}\label{ldpcresul1}
\end{figure}

\begin{figure}[!h]
\centering
\includegraphics[width=9cm]{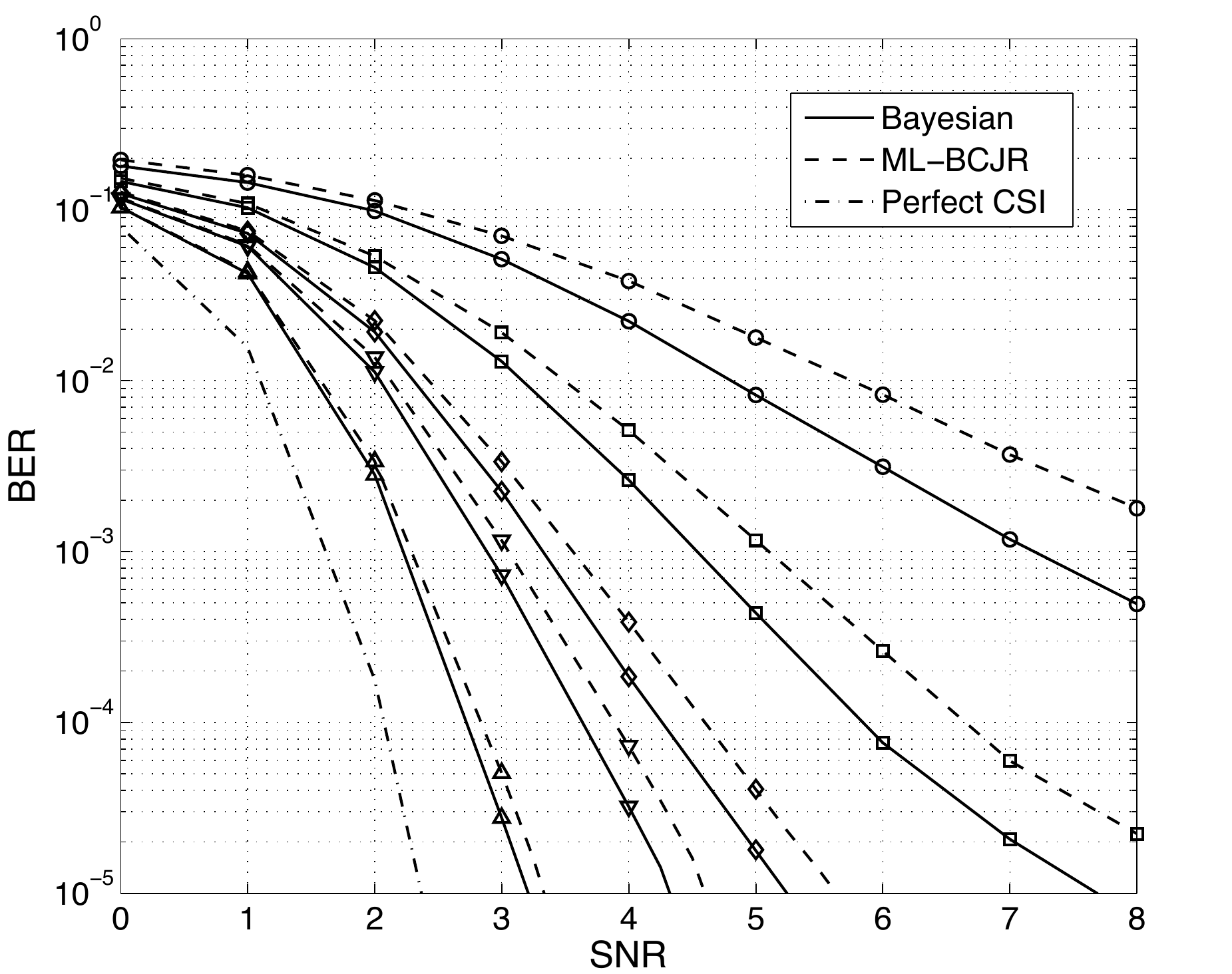}
\caption{BER performance for Bayesian equalizer (solid lines) and ML-BCJR (dashed lines) after the decoder, for a channel with 6 taps and different lengths of the training sequence, $n=15$ ($\circ$), $n=25$ ($\square$), $n=40$ ($\diamond$), $n=50$ ($\bigtriangledown$) and $n=90$ ($\bigtriangleup$). Dash-dotted line illustrates performance assuming perfect CSI.}\label{ldpcresul6taps}
\end{figure}

In Fig. \ref{ldpcresul1} and \ref{ldpcresul6taps} we show the BER after the LDPC decoder has corrected the errors introduced by the channel. In these plots, we can observe that for short training sequences the gains of using the Bayesian equalizer over the ML-BCJR are significant (over $1$ dB). The difference between the results in Fig. \ref{ldpcresul6tapsdetect}, and Fig. \ref{ldpcresul1} and \ref{ldpcresul6taps} can be explained by the APP estimates given by each procedure. When we measure the BER after the equalizer (Fig. \ref{ldpcresul6tapsdetect}), we only care if each bit has been correctly decoded and, consequently, we are only measuring how good the APP estimate of the 50\% percentile is. When we measure the BER after the LDPC decoder (Fig. \ref{ldpcresul1} and \ref{ldpcresul6taps}), we need that the APP estimates are accurate everywhere, because the LDPC decoder needs these estimates to decode correctly the transmitted codeword, i.e., the Belief Propagation algorithm uses the APP for each individual bit. These results sustain our claim that the Bayesian equalizer provides more accurate predictions of the APP than the ML-BCJR equalizer, as the LDPC decoder is able to better decode with them.

Particularly, in Fig. \ref{ldpcresul1} we can observe that the Bayesian equalizer improves the ML-BCJR equalizer as we increase the SNR for a fixed length of the training sequence and both equalizers tend to coincide as we increase the length of the training sequence. This last result is expected, because as we increase the length of the training sequence the posterior for the CSI tends to a delta function and the ML-BCJR and Bayesian equalizers coincides. But for these equalizers to coincide, we need over 60 training samples (more than 20 per coefficient) and for shorter training sequences the Bayesian equalizer is vastly superior. 

Finally, in Fig. \ref{ldpcresul6taps}, we plot the curves for the channel with 6 taps. In this experiment, we can see the previous result magnified, because the longer the impulsive response of the channel is the more uncertain the channel estimate is and the higher the room for improvement for the Bayesian equalizer. In this experiment, we can observe gains over $1$ dB for training sequence with 25 symbols and $\mbox{SNR}=7$ dB.

\section{Conclusions and Future Work}\label{sec5}
Channel equalization has been traditionally solved using a discriminative model where the only variable modeled as random is the noise. The generative model introduced in this paper, where the posterior probability density function of the estimated CSI is included, is a more efficient solution. If we are to just estimate the encoded transmitted symbols, the discriminative model is a good choice. However, if the estimation of the APP is needed, i.e., the decoder very much benefits from this information, the discriminative solution exhibits poor results whenever the CSI is badly estimated. On the contrary, the Bayesian approach exploits the full statistical model to provide optimal APP. We prove that these estimations are useful if a LDPC encoding is used. Other decoders may take advantage of this solution as well.

Despite both the BCJR algorithm and LDPC decoding can be computed efficiently using machine learning algorithms applied on graphs \cite{Frey01}, the drawback of this proposal is its computational complexity, because we have to compute several times the results for the BCJR algorithm to average the integral (\ref{bayesint}). Alternative graphical representation or inference algorithms that capture the essence of this approach  may yield sub-optimal but less demanding solution for Bayesian equalization proposed. Furthermore, other systems and channels can be considered.
width=2.5in]{subfigcase1}

\section*{Acknowledgment}
This work was partially funded by Spanish government (Ministerio de Educaci\'on y Ciencia TEC2006-13514-C02-\{01,02\}/TCM and TEC2009-14504-C02-\{01,02\}, Consolider-Ingenio 2010 CSD2008-00010), and the European Union (FEDER).

\bibliographystyle{IEEEtran}

\bibliography{LDPC1,gp1,macler,UMTS1,digwircom_short,PredisNLEq,svm,eq1}

\end{document}